%% file: yojul1.tex
\documentstyle[twocolumn,seceq]{jpsj}

\def\runtitle{Phase Diagram and Pairing Symmetry of the 2D $t$-$J$ Model}
\def\runauthor{Hisatoshi {\sc Yokoyama} and Masao {\sc Ogata}}

\title
{
Phase Diagram and Pairing Symmetry of the Two-Dimensional $t$-$J$
Model by a Variation Theory
}

\author
{
Hisatoshi {\sc Yokoyama}\footnote{Electronic address:
yoko@cmpt01.phys.tohoku.ac.jp}
and Masao {\sc Ogata}$^{1,}$\footnote{Electronic address:
ogata@sola.c.u-tokyo.ac.jp}
}

\inst
{
Department of Physics, Tohoku University,
Aramaki Aoba, Aoba-ku, Sendai 980-77\\
$^1$Department of Basic Science, Graduate School of Arts and Sciences,
University of Tokyo, Komaba, Meguro-ku, Tokyo 153
}

\recdate
{
\qquad\qquad\qquad\qquad\qquad
}

\abst
{
Two-dimensional $t$-$J$ model is studied by a variational
Monte Carlo method, using Gutzwiller-Jastrow-type
wave functions.
Various kinds of superconducting pairing symmetries are
compared in order to determine the phase diagram of the ground state in the
full $J/t$-$n$ plane.
Near the half filling where the
high temperature superconductivity is expected,
the d$_{x^2-y^2}$ wave pairing state is always the most stable among various
symmetries.
The three-site term hardly changes the phase diagram in this regime.
In the low electron density, the extended s-type wave becomes a quantitatively
good state for large $J/t$
although the energy gain is small.
The Gutzwiller wave function is shown to be the exact
ground state in the low-electron-density limit for the supersymmetric
case ($J/t=2$).
}

\kword
{
high temperature superconductivity, $t$-$J$ model,
variation theory, pairing symmetry, three-site term,
Monte Carlo method, phase diagram, Gutzwiller wave function,
Gutzwiller approximation
}

\begin{document}
\sloppy
\maketitle

\input paperfm
\section{Introduction}
Symmetry of Cooper pairs is considered as a key point
to elucidate the mechanism of high-$T_{\rm c}$ superconductivity.
Many extensive experiments suggest that the unconventional
d$_{x^2-y^2}$ anisotropic pairing is realized in high-$T_{\rm c}$
oxides.\cite{Scalapino}
Interesting phenomena due to this anisotropic pairing have been
studied phenomenologically.
Actually the two-dimensional (2D) $t$-$J$ model, which is considered
as an effective Hamiltonian for the high-$T_c$ oxides,\cite{RVB}
will probably have a ground-state with the d$_{x^2-y^2}$-wave
pairing state for the reasonable parameters.
\par

The 2D $t$-$J$ model is derived by regarding a singlet
between the Cu spin and the hole's spin
on the neighboring O sites (Zhang-Rice singlet)
as a mobile vacancy in the Heisenberg spin system.\cite{ZR}
Different approaches, such as mean-field
theories,\cite{BZA,Kotliar,KL,Suzu,ZGRS,Tanamoto}
variational studies\cite{YSS,Gros,GL} and exact diagonalization
methods,\cite{Dagotto,Ohta,Poilblanc} support
the d$_{x^2-y^2}$-wave pairing near half filling and for $J/t \sim 0.4$.
\par

In this paper we study more extensively various pairing states
in the whole parameter space ($J/t$ and electron density $n$)
using a variational Monte Carlo method,
and determine the phase diagram of the ground state.
\par

It seems natural that the strong on-site repulsion
between up-spin and down-spin electrons favors
the anisotropic d$_{x^2-y^2}$-wave pairing.
However, it is not at all apparent why, for example, the extended s-wave
pairing state, which does not have the on-site pairing amplitude either,
is not realized.
\par

Furthermore, many discussions on the d$_{x^2-y^2}$-wave pairing
have been limited to the low doping regime
(near the half filling) and to small values of $J/t$.
If we consider the wider range of parameter space,
it is necessary to clarify the region where
the d$_{x^2-y^2}$-wave is stable.
For example, in the low-electron-density side a metallic state
is expected, while in the large-$J/t$ region a phase separation is
predicted.~\cite{Emery,Putikka1}
The relation between the phase separation and the d$_{x^2-y^2}$-wave
state is not so clear.
Actually from the phase diagram of the one-dimensional (1D) $t$-$J$ model,
we expect that the superconducting state is next to the
phase separation.~\cite{Ogata1d}
But even if this is the case in 2D, we have to clarify the
pairing symmetry of the neighboring superconducting state.

Meanwhile, in the low-electron densities, there may be a
region where the s-wave pairing becomes stable.\cite{Dagotto}
The two-electron problem is analytically
solved,\cite{Emery,Lin,Kagan,HManousakis}
to show a stable s-wave bound state for $J/t\ge 2$.
This suggests that the extended s-wave pairing state is favorable
for $J/t\ge 2$ in the low-density region.  
Actually in the 1D $t$-$J$ model, such a state was expected.\cite{Ogata1d}
For finite electron densities in the 2D case, we have to check
this possibility.
It is far from clear how this s-wave state relates to the
d$_{x^2-y^2}$-wave state as well as to the phase separation.
\par

For the supersymmetric case ($J/t=2$), or on the edge of
this extended s-wave region, we expect that the Gutzwiller
wave function (GWF)\cite{Gutz} becomes the exact ground state
for $n\rightarrow 0$.
This fact was predicted before in connection with the 1D $t$-$J$
model.\cite{YO1}
It is very interesting to clarify the role of the GWF in the phase
diagram.
\par

With these aspects in mind, we focus on the following points
in this paper.
(1) How and why is the d$_{x^2-y^2}$-wave pairing stabilized,
but not the other pairing states for the high-$T_{\rm c}$ parameters?
(2) The ranges of the stable d$_{x^2-y^2}$-wave state as well as
other ones.  Namely is there any possibility for the other pairing
state to be realized in the phase diagram?
(3) How large is the region where the phase separation takes place?
(4) Is there any region where the extended s-wave state is stable
in the low-electron density?
(5) How are the variational states connected to the states
in the low electron density limit and what happens at the
supersymmetric case ($J/t=2$)?
(6) How does the three-site term affect the above results?
\par

In order to study these questions on the phase diagram, we use
the variational Monte Carlo (VMC) method.
So far, the exact treatments like the quantum Monte Carlo methods
have been often useless for the 2D $t$-$J$ model because of the
minus sign problem.
On the other hand, the present VMC method is free from such an
obstacle and treats exactly the constraint of no doubly occupied
sites.\cite{YSS,Gros,GL}
Accurate estimates of the expectation values are possible,
which are indispensable to take advantage of merits of
the variation theory.
\par

As variational wave functions for superconductivity, we use
Gutzwiller-projected BCS functions which were originally proposed
by Anderson as a resonating valence bond state.\cite{RVB}
It has been shown that the d$_{x^2-y^2}$-wave pairing state is the
best variational state among Gutzwiller-projected BCS states
near half filling.\cite{YSS,Gros,GL}
In this paper, we extend previous works to include a variety of pairing
symmetries in the full parameter space of the 2D $t$-$J$ model
with the three-site term.
We construct a phase diagram of the ground state within
the Gutzwiller-Jastrow-type wave functions.
As will be seen later, the resultant 2D phase diagram shares
some features with the 1D phase diagram.\cite{Ogata1d}
\par

The mean-field theories are sometimes useful to predict the
behaviors of the system
qualitatively.\cite{BZA,Kotliar,KL,Suzu,ZGRS,Tanamoto}
However, they treat the constraint of no double occupancy
approximately.
We are going to compare the obtained phase diagram
with that predicted in the mean-field theories with the Gutzwiller
approximation.
It is found that a qualitatively correct phase diagram is obtained,
but quantitatively we need some modifications
for the Gutzwiller approximation.
\par

The outline of this paper is as follows:
In \S2 we explain the model Hamiltonian and the variational
wave functions.
In \S3 the relation between our variational states and the slave-boson
mean-field theory is mentioned.
Section 4 is devoted to the discussion on the metallic,
magnetic and superconducting states for the high-electron-density
or low-doping regime.
We study the properties of the stable d$_{x^2-y^2}$-wave state
and search for the origin of its stability in terms of the
resonating valence bond theory.
In \S5 a phase diagram is constructed in the full
$J/t$-$n$ plane, and properties of wave functions
are described in the low electron density.
We summarize in \S6.
In Appendix A the energy of the Hubbard model is shown
for a comparison in \S4.1.
The brief summary of the Gutzwiller approximation is given in
Appendix B.
\par

A part of the present results was published before.~\cite{YKO}
\par

\setcounter{equation}{0}
\section{Model and Trial Wave Functions}
We consider the $t$-$J$ model defined in the Hilbert space
without double occupancy of electrons.
In this paper we include the so-called ``three-site term"
or ``pair-hopping term", ${\cal H}_3$, which is usually neglected
but is present in the effective Hamiltonian derived from the
Hubbard model.\cite{Harris}
Thus our model is written as
\begin{equation}
{\cal H}={\cal H}_t+{\cal H}_J+{\cal H}_3,
\end{equation}
\begin{equation}
{\cal H}_t=-t\sum_{\langle i,j\rangle\sigma}
\Bigl(c^\dagger_{i \sigma}c_{j \sigma}+c^\dagger_{j \sigma}c_{i \sigma}\Bigr),
\end{equation}
\begin{equation}
{\cal H}_J=J\sum_{\langle i,j\rangle}\left({\mib S}_i \cdot{\mib S}_j
          -\frac{1}{4}n_in_j\right) ,
\end{equation}
\begin{eqnarray}
{\cal H}_3=-\frac{J^{(3)}}{4}&&\sum_{j,\tau\ne\tau',\sigma}
           \left(c^\dagger_{j,-\sigma}c_{j,-\sigma}
                 c^\dagger_{j+\tau,\sigma}c_{j+\tau',\sigma} \right. \nonumber \\
                 &&\left.+c^\dagger_{j+\tau,-\sigma}c_{j,-\sigma}
                 c^\dagger_{j,\sigma}c_{j+\tau',\sigma}\right),
\end{eqnarray}
where 
$\langle i,j\rangle$ represents the sum over the nearest-neighbor
pairs.
$\tau$ and $\tau'$ run as
vectors pointing to the nearest-neighbor sites.
We take $t$ as the energy unit in the following.
This model has four parameters: $J/t$, $J^{(3)}/J$, electron density
$n=N/N_{\rm a}$ and magnetization $m=(N_\uparrow-N_\downarrow)/N_{\rm a}$,
where $N$ ($=N_\uparrow+N_\downarrow$) is the total electron number,
$N_{\rm a}$ the number of sites and $N_\sigma$ the number of electrons
with spin $\sigma$.
We study mostly the nonmagnetic case ($m=0$).
\par

The three-site term, ${\cal H}_3$, shows up in the
effective Hamiltonian of the Hubbard model in the strong coupling
regime ($U\gg t$).~\cite{Harris}
In this case $J^{(3)}$ is equal to $4t^2/U(=J)$.
For the electron-doped high-$T_c$ cuprate (Nd system),
the doped electrons directly
enter the Cu orbitals, so that the Hubbard model becomes an adequate
model\cite{Nd} and thus we should assume $J^{(3)}=J$.
\par

On the other hand,
if the effective Hamiltonian is derived from a hole-doped
CuO$_2$ model including Cu d$_{x^2-y^2}$ and O 2p orbitals,~\cite{ZR,RP}
$J^{(3)}$ could be negative.~\cite{BA,Matsukawa}
Therefore in this paper we use $J^{(3)}$ as a parameter and discuss
its effect on the phase diagram.

\par
For this Hamiltonian, we study Gutzwiller-Jastrow-type variational
wave functions defined as:
\begin{equation}
\Psi=\prod_{j,\ell}\prod_{\sigma\sigma'}
\Bigl[1-\Bigl\{1-\eta(r_{j\ell})\Bigr\}n_{j\sigma}n_{\ell\sigma'}\Bigr]\Phi,
\end{equation}
with $r_{j\ell}=|{\mib r}_j-{\mib r}_\ell|$.
The function $\Phi$ is a one-body mean-field-type wave function;
we consider (1) a simple Fermi sea
$\Phi_{\rm F}$ as a metallic state, (2) a Hartree-Fock
antiferromagnetic (AF) function
for the AF ordered state, and (3) a BCS wave function
with various pairing symmetries.
So far these types of wave functions have been widely used for the
small-doping region.
In this paper, we study these states in the whole parameter regime.
\par

The AF state state is defined as\cite{Ogawa,YS2,YS2tj}
\begin{equation}
\Phi_{\rm AF}=\prod_{k,\sigma}\left[{\tilde u}_k\ c_{k\sigma}^\dagger
    +{\rm sgn}(\sigma)\ {\tilde v}_k\ c_{k+K\sigma}^\dagger\right]|0\rangle,
\end{equation}
with
$$
\tilde u_k=\left[\frac{1}{2}\left(1-
    \frac{\varepsilon_k}{\sqrt{\varepsilon_k^2
                                   +\Delta_{\rm AF}^2}}\right)\right]^{1/2},
$$
$$
\tilde v_k=\left[\frac{1}{2}\left(1+
    \frac{\varepsilon_k}{\sqrt{\varepsilon_k^2
                                   +\Delta_{\rm AF}^2}}\right)\right]^{1/2}.
$$
Here, sgn$(\sigma)$ takes either $+1$ or $-1$ according as
$\sigma=\uparrow$ or $\downarrow$; $\varepsilon_k=-2t(\cos k_x+\cos k_y)$
and the wave vector
$K=(\pi,\pi)$ represents the AF reciprocal lattice vector.
$\Delta_{\rm AF}$ is an AF variational parameter;
$\Phi_{\rm AF}$ is reduced to the simple Fermi sea $\Phi_{\rm F}$
as $\Delta_{\rm AF}\rightarrow 0$.
\par

The BCS wave function is given as\cite{YSS,Gros,GL}
\begin{equation}
\Phi_{\rm SC}=
\prod_k
\left[u_k+v_kc^\dagger_{k\uparrow}c^\dagger_{-k\downarrow}\right]|0\rangle,
\end{equation}
with
\begin{equation}
a_k=\frac{v_k}{u_k}=\frac{\Delta_k}
{{\varepsilon_k-\mu}+\sqrt{(\varepsilon_k-\mu)^2+\Delta_k^2}}.
\end{equation}
For most cases
we take $\mu$ as the chemical potential for the noninteracting
system, $\mu_0$.
However the optimal value of $\mu$ ought to change due
to the correlation.  We will check
the $\mu$ dependence of the variational energy in \S4.3.
\par

The pairing symmetry of $\Delta_k$ is chosen as follows:
\begin{equation}
\Delta_k=\left\{
\begin{array}{ll}
\Delta & \mbox{\quad(i)\quad s wave}\\
\Delta\sin k_x\sin k_y & \mbox{\quad(ii)\quad d$_{xy}$ wave}\\
\Delta(\cos k_x+C\cos k_y) & \mbox{\quad(iii)\quad s+d wave}\\
\Delta(\cos k_x+e^{i\theta}\cos k_y) & \mbox{\quad(iv)\quad
                                             s+$i$d wave}.
\end{array}
\right.
\end{equation}
In each case $\Delta\ (\ge 0)$ is a variational parameter and
when $\Delta=0$, $\Phi_{\rm SC}$ is reduced to the simple
Fermi sea, $\Phi_{\rm F}$.
Note here that the expectation value of
the order parameter 
$\langle c_{k\uparrow}^\dagger c_{-k\downarrow}^\dagger \rangle =
\Delta_{\rm SC}$
is not necessarily equal to $\Delta$.
$\Delta_{\rm SC}$ is to be calculated as an expectation
value in the variational state.
Actually $\Delta_{\rm SC}$ is largely reduced from $\Delta$
near the half filling
due to the strong electron correlation.\cite{YSS,ZGRS}
\par

By changing the value of $C$ ($-1\le C\le 1$) and $\theta$
($0\le\theta\le\pi$) in the cases (iii) and (iv),
various symmetries are continuously realized.
For example, $\Delta_k$ of the case (iii) is rewritten as:
\begin{equation}
\Delta_k=\Delta_{\rm s}(\cos k_x+\cos k_y)
        +\Delta_{\rm d}(\cos k_x-\cos k_y),
\end{equation}
with $\Delta_{\rm s}=\Delta(1+C)/2$ and $\Delta_{\rm d}=\Delta(1-C)/2$.
Thus the state with $C=-1$ represents the pure d$_{x^2-y^2}$ symmetry, the
state with $C=0$ s+d state, and the state with $C=1$ the pure extended s
state, respectively.
For the case $-1<C<1$, the wave function has different gap
amplitudes in the directions of $x$ and $y$ axes, which
conflicts with the symmetry of the lattice.
On the other hand, $\Delta_k$ of the state (iv) represents a complex
mixture of the d$_{x^2-y^2}$ and extended s states.
Especially at $\theta=\pi/2$, $\Delta_k$ is the so-called s+$i$d
state.\cite{Kotliar}
\par

Many-body effects are introduced into eq.~(2.5) by a
two-body Jastrow correlation factor
\begin{equation}
{\cal P}=\prod_{j,\ell}\prod_{\sigma\sigma'}
\Bigl[1-\Bigl\{1-\eta(r_{j\ell})\Bigr\}n_{j\sigma}n_{\ell\sigma'}\Bigr].
\end{equation}
This factor reduces the amplitude of the wave function according
to the charge configuration.
By requiring $\eta(0)=0$, the local constraint of the prohibition
of double occupancy (Gutzwiller projection) is strictly satisfied.
For $r\ne 0$, we consider two cases: (a) $\eta(r)=1$, which is
nothing but the Gutzwiller projection
${\cal P}_{\rm d}=\prod_j[1-n_{j,\uparrow}n_{j,\downarrow}]$ and
(b) an intersite correlation factor,
\begin{equation}
\eta(r)=\left[\frac{L}{\pi}\sqrt{\sin^2\left(\frac{\pi x}{L}\right)
 +\sin^2\left(\frac{\pi y}{L}\right)}\right]^\nu.
\end{equation}
Here $L$ is the linear dimension of the lattice: $N_{\rm a}=L\times L$.
We abbreviate this correlation factor as ${\cal P}_{\rm TLL}$.
It works as repulsive Jastrow correlation for $\nu>0$ and
attractive one for $\nu<0$, and is reduced to
${\cal P}_{\rm d}$ for $\nu=0$.
This factor is a Tomonaga-Luttinger-liquid (TLL) type
and has been investigated in detail for the 1D,\cite{HM,YO1} and
the 2D systems.\cite{VG}
Here we choose this form only for its simplicity as an intersite
factor.  
Alternatively one can use a short-range correlation factor,\cite{YO1,YS3}
but the difference of the optimized variational energy
between the two choices is negligibly small.
\par

\setcounter{equation}{0}
\section{Relation between Variation Theory and Slave-Boson
Mean-Field Approximation}

The phase diagram and physical properties of the 2D $t$-$J$ model
were extensively studied in the slave-boson mean-field
approximation.~\cite{Tanamoto}
Thus it is important to clarify the relation to the present
variation theory.
In the slave-boson mean-field approximation, the electron
operator at site $i$ and spin $\sigma$ is decoupled as\cite{SB}
\begin{equation}
c_{i,\sigma} = b_i^\dagger f_{i,\sigma},
\end{equation}
where $b_i$ is a slave-boson operator which represents the
vacant site (holon), and $f_{i,\sigma}$ is a fermion (spinon)
operator.
\par

The mean-field approximation using this slave-boson representation
gives a d-wave pairing state as the ground state near
half filling.\cite{Kotliar,KL,Suzu}
In this state, the fermion (spinon) degrees of freedom have a BCS
order parameter
\begin{equation}
\langle f_{i,\uparrow}^\dagger f_{i+x,\downarrow}^\dagger \rangle
= - \langle f_{i,\uparrow}^\dagger f_{i+y,\downarrow}^\dagger \rangle,
\end{equation}
and its wave function is the same with eq.~(2.7).
\par

On the other hand,
the slave bosons are assumed to form a bose condensate.\cite{KL,Suzu}
Since the slave bosons have hard-core nature or two bosons
cannot occupy the same site,
it is not clear whether the bosons really form a bose condensate
in the two (or quasi-two) dimensional systems.
In the slave-boson mean-field approximation, however, the hard-core
nature is neglected and thus the ground-state for the bosons is written as
\begin{equation}
(b_{k=0}^\dagger)^{N_b} |0\rangle,
\end{equation}
where $N_b$ is the number of holes.
\par

We can take account of the hard-core nature by using a
Gutzwiller projection operator for bosons
\begin{equation}
{\cal P}_b (b_{k=0}^\dagger)^{N_b} |0\rangle,
\end{equation}
instead of (3.3).  Even in this case, the wave-function amplitude
for the slave boson is equal to 1 for any boson configurations.

The electronic wave-function in the slave-boson scheme is
the product of the fermion and boson wave functions.
Therefore the ground-state
obtained in the slave-boson mean-field theory
is equal to eq.~(2.7);
our variational calculations take full account of the effect of
the Gutzwiller projection.
\par

The VMC method\cite{YKO,MacCep} enables us to estimate accurate
expectation values even in the presence of the Gutzwiller projection.
We use square lattices with $L=10$-26 for the metallic
states and $L=10$-20 for the ordered states; mainly the system
with $L=10$ is used for $\Psi_{\rm SC}$, because the size dependence
is negligible for $L\ge10$ as far as energy is concerned.
The periodic($x$)-antiperiodic($y$) boundary conditions are used
and $N_\sigma$ is chosen so as to satisfy the closed-shell
condition.
To suppress statistical errors, we take sufficient sample
numbers ($3\times10^4$ - $2\times10^5$) and Metropolis trials
(50 at maximum for each electron) between the samplings.
\par

To treat the coherent states for the superconductivity, two
VMC schemes have been applied: one by using the grand canonical
scheme~\cite{YSS} and one by fixing the electron
number.~\cite{Gros,Lhuillier}
The results of the two methods ought to coincide
in the thermodynamic limit.
The former scheme has a merit in calculating the order
parameter directly, but it needs much longer CPU time.
In this work we adopt the latter scheme.
\par

\setcounter{equation}{0}
\section{Pairing Symmetry in Low-Doping Regime}

In this section we study the region of high electron density
or low doping, laying emphasis on the stable pairing symmetry of
superconductivity.
In \S4.1 we study metallic states as a starting point, since the
other states with AF order or superconducting order parameter
reduces to the GWF when the order parameter vanishes.
The energy gain due to the occurrence of the order is discussed
in comparison with the variational energy of the GWF.
The instability of metallic states against the phase
separation and the effect of the Jastrow correlation factor are
described.
Next we investigate various pairing states for the half-filled case
in \S4.2 and for less-than-half-filled cases in \S4.3.
We show that the ordered states are free from the instability
against the phase separation for a small value of $J/t$.
Finally in \S4.4 we look at the properties of the stable
d$_{x^2-y^2}$-wave state.
\par

\subsection{Metallic states}

First, we consider the GWF,
$\Psi_{\rm G}={\cal P}_{\rm d}\Phi_{\rm F}$, as a metallic
state.
In Fig.~1 three energy components:
transfer energy $E_t=\langle{\cal H}_t\rangle$,
exchange energy $E_J=\langle{\cal H}_J\rangle$ and
the three-site contribution $E_3=\langle{\cal H}_3\rangle$
are plotted as a function of $n$;
size dependence is almost negligible in this scale.
\par

Near the half filling, $E_J$ is upward convex.
This means that when $J/t\rightarrow\infty$ the
system must have a phase separation.
In order to study this kind of phase separation in the small-$J/t$
region, it is necessary to study the $n$ dependence of
variational energies more carefully.
Figure 2 shows the close-up of this dependence near
the half filling.
On one hand, $E_t/t$ and $E_3/J^{(3)}$ are linear with respect to
$\delta\equiv 1-n$; on the other hand,
$E_J/J$ does not behave linearly but
\begin{equation}
E_J/J=E_J(\delta=0)/J+{\rm const.}\times\delta^{0.7},
\end{equation}
for wide range of $n$.\cite{note0}
This $\delta$-dependence means that the total energy $E=E_t+E_J+E_3$
is always upward convex for any non-zero value of $J/t$.
Thus the GWF is unstable in itself against the phase separation.
\par

As shown in Appendix A, this is in sharp contrast with the GWF in
the Hubbard model, for which both $E_t/t$ and $E_U/U$ are proportional
to $\delta$ as $\delta\rightarrow 0$.
The total energy is always downward convex, leading to
no phase separation.
This indicates that as long as the 2D GWF is concerned,
the $t$-$J$ model for $J/t>0$ is not connected to the
Hubbard model near the half filling.
\par

We study this instability also for the TLL-type function,
$\Psi_{\rm TLL}={\cal P}_{\rm TLL}\Phi_{\rm F}$,
which Valent\'\i\ and Gros studied.~\cite{VG}
The results are summarized in Table I, indicated as TLL.
As $n\rightarrow 1$ (i.e.\ $\delta \rightarrow 0$),
intersite correlation factors generally
become less effective, so that the occurrence of the phase
separation near the half filling is not affected even by
this intersite correlation factor.
Namely, as far as a form of ${\cal P}\Phi_{\rm F}$ is assumed,
the instability against the phase separation is inevitable.
Later when we study the projected BCS state, this phase separation
is removed.
\par\medskip

\subsection{Half-Filled Case}

First, let us look at the results for the half filling.
Figure 3 shows the variational energy for the the case (iii)
in eq.~(2.9) for various values of $C$.
For the d$_{x^2-y^2}$-wave case ($C=-1.0$), $E_J/J$ is symmetric
with respect to $\Delta/t=2$ and thus has two minima at
$\Delta/t\sim 0.57$ and 7.
(This fact was not discussed in the previous VMC calculations.~\cite{Gros})
As the parameter $C$ increases, the minimal value monotonically
increases, as seen in the inset, where the scale is expanded.
Thus the d$_{x^2-y^2}$ wave has the lowest energy in the class
of variational states (iii).
For the extended s wave ($C=1.0$) there is no dependence
on $\Delta$ of the variational energy.\cite{Gros}
This is because $\mu=0$ at half filling and thus $a_k$ becomes a
constant, which means that $\Psi_{\rm SC}$ is equivalent
to $\Psi_{\rm G}$.
\par

In Fig.~4 we show the variational energy for
the s+$i$d-type wave function (iv) in eq.~(2.9).
As in the inset the
energy minima for $\pi/2\le\theta\le\pi$ are degenerate within
the error bars.
This degeneracy is due to the SU(2) symmetry of the Heisenberg
model;\cite{Affleck} it was proved that under the Gutzwiller
projection the state with $\pi/2\le\theta\le\pi$ are all
the same state.\cite{ZGRS}
\par

These VMC calculations for the projected BCS-type wave functions
$\Psi_{\rm SC}$ are consistent with the slave-boson mean-field
theory.
However, the actual ground state at half filling should be an
AF magnetically ordered state.
In fact, $\Psi_{\rm AF}={\cal P}_{\rm d}\Phi_{\rm AF}$
is slightly stabler than $\Psi_{\rm SC}$ and
the best variational state among the present states.
This feature is in contrast with 1D systems, where
ordered states---AF~\cite{YS2tj} and s-type superconducting~\cite{Y}
states--- have higher energies than the GWF.
The expectation values of $\langle {\mib S}_i\cdot{\mib S}_j\rangle$
for the nearest-neighbor $i$ and $j$ are $-0.3206$ for
$\Psi_{\rm AF}$,\cite{YS2tj} and $-0.3199$ for the d$_{x^2-y^2}$
or the s+$i$d state, which are close to the best
estimation,\cite{TC} $-0.3346$.
\par\medskip

\subsection{Symmetry of Cooper pairs in less-than-half-filling}

Now, we consider less-than-half-filled cases.
We concentrate on the s+$i$d-type wave function, (iv) of eq.~(2.9),
which contains the d$_{x^2-y^2}$ and extended s-wave states.
All the other states give higher energy than the s+$i$d-type state.
For example, the s+d-type-wave states (iii) in eq.~(2.9)
with $-1<C<1$ always give higher variational energy
as in the half-filled case.
As for the d$_{xy}$-wave state (ii), we find that each
energy component is a monotonically increasing
function of $\Delta/t$ for every electron density.
Thus there is no chance for the
d$_{xy}$ wave to be stabilized.
The behavior of the conventional s-wave state (i) is very similar to
the extended s wave in the high density regime; the energy
is always higher than the d$_{x^2-y^2}$ wave for small $\delta$.
\par

Let us return to the s+$i$d-type state.
Figures 5(a)-(c) show the energy components of the s+$i$d-type
wave function.
As a typical electron density, we use $\delta=0.12$ ($n=0.88$).
The d$_{x^2-y^2}$ wave ($\theta=\pi$) is the most favorable except
the three-site term.
Thus, when there is no three-site term ($J^{(3)}=0$),
the d$_{x^2-y^2}$-wave state has always the lowest variational energy.\cite{note1}
As an example, we show in Fig.~6 the total energy $E/t$
for $J/t=0.25$ and several values of $\theta$.
The situation does not change for $J/t\lsim 2.0$.
\par

Here we mention the effect of the three-site term ${\cal H}_3$.
The variational energy for the extended s wave is lowered in the
presence of ${\cal H}_3$.
We have found, however, that the most stable state is still the d$_{x^2-y^2}$
wave, even when the value of $J^{(3)}/J$ is as large as 1.0.
On the contrary, the d$_{x^2-y^2}$ wave is further stabilized
in the ferromagnetic case ($J^{(3)}<0$).
The situation does not change for any value of $J/t$ and
for electron densities $n=0.52$-0.96.
\par

Figure 7 shows the optimal variational energy for the d$_{x^2-y^2}$-wave
state for $n\sim 1$ and $J/t=0.5$.
In comparison with the metallic GWF, there is a definite energy gain
due to the superconducting order parameter.
In Fig.~7 we also plot the data of the AF state,
$\Psi_{\rm AF}={\cal P}_{\rm d}\Phi_{\rm AF}$
and the $\pi$-flux state.
Just at half filling $\Psi_{\rm AF}$ is slightly lower than
the d$_{x^2-y^2}$-wave state.
However even at $\delta=0.02$ the d$_{x^2-y^2}$-wave state
has a lower energy, so that the AF state disappears immediately
when holes are doped in the 2D Heisenberg system.
\par

The $\pi$-flux state,\cite{Kotliar} which is equivalent to the
s+$i$d state at half filling, has exactly the same energy with
the d$_{x^2-y^2}$-wave state for $n=1$.\cite{KL,ZGRS,Affleck}
For the doped case, however, the $\pi$-flux state has much higher
variational energy than the d$_{x^2-y^2}$ wave.
It is even higher than the AF state.  Therefore it is not likely
that the d$_{x^2-y^2}$-wave state is mixed with the $\pi$-flux state.
\par

Finally, we refer to the $\mu$ dependence of the energy.
We calculate the total energy for some $J/t$ by varying
the value of $\mu$ in eq.~(2.8). 
The smaller $J/t$ is, the more weakly $E/t$ depends on $\mu$.
As an example, the result for $J/t=0.5$ is plotted in
Fig.~8, where $\Delta_{\rm opt}/t \sim 0.55$.
The energy improvement by optimal $\mu$ upon the value
for $\mu_0$ is slight.  
Thus the use of the noninteracting value $\mu_0$ is
justified for small $J/t$ to some extent, although this
is not the case for a value of $J/t$ as large as 2.
\par

\subsection{Properties of d$_{x^2-y^2}$ wave}
In this subsection we compare the properties of the d$_{x^2-y^2}$-wave
state with other variational states in order to
consider the reason of its stability.
In Table I, we compare each energy component for some variational
states; the optimal functions are used for the TLL, AF and
d$_{x^2-y^2}$-wave states.
The differences from the GWF value are shown.
\par

First of all, we notice that the metallic state with Jastrow
correlation (indicated as TLL) has slightly lower kinetic energy
($<1$\% ) than the GWF.
This is because the Jastrow (repulsive) correlation factor tends
to separate the electrons so as to gain the kinetic energy.
However its gain is so small that TLL state cannot overcome the
d$_{x^2-y^2}$-wave state.

In contrast to this, the other ordered states (AF, d, s+$i$d state)
unanimously gain the exchange energy, namely
$\langle {\mib S}_i\cdot{\mib S}_j\rangle$ and
$\langle -n_in_j/4\rangle$, but all of them lose the kinetic energy.
As the doping increases the energy gain for the exchange term
becomes smaller and smaller.
Finally at some doping, the energy gain due to the AF order or
superconducting order vanishes.
This tendency gives the $n$-dependences of the variational energy
shown in Fig.~7.

Comparing the d$_{x^2-y^2}$-wave with the AF states, we can see that
the loss of the kinetic energy is always smaller than that for
the AF state, while the gain of the exchange energy is larger for
the d$_{x^2-y^2}$ wave for every electron density.
This makes the d$_{x^2-y^2}$ the best variational state.
\par

In Table I, we also include the energy components for the
s+$i$d and extended s states.
For comparison we borrow the same value of $\Delta/t$ with
the optimal d$_{x^2-y^2}$ case, namely $\Delta/t=0.7, 0.55, 0.40$,
and $0.25$ for $n=096, 0.88, 0.80$, and $0.72$, respectively.
As for the exchange term
$\langle {\mib S}_i\cdot{\mib S}_j\rangle$, the d$_{x^2-y^2}$-wave
state has a large energy gain.
For example at $n=0.88$ ($\delta = 0.12$)
it gains 29\% of the exchange term of the GWF.
The s+$i$d-wave also gains the exchange term similar to the
d$_{x^2-y^2}$ wave,
but its magnitude is slightly smaller.
For the extended s wave the energy gain is very small.
This is the reason why the d$_{x^2-y^2}$-wave state is favored by the exchange term $J$.
We discuss this phenomena in terms of the resonating valence bond
(RVB) representation of the wave functions.

Equation (2.7) can be rewritten as
\begin{equation}
\Psi_{\rm SC}
={\cal P}_d \prod_k u_k \cdot {\rm exp}
\biggl[ \sum_{i,j}
a_{i,j} c^\dagger_{i\uparrow}c^\dagger_{j\downarrow}\biggr] |0\rangle,
\end{equation}
where $a_{i,j}$ is Fourier transform of $a_k$.
If we fix the electron number by projecting out the $N$ electron states,
we obtain
\begin{equation}
{\cal P}_N \Psi_{\rm SC}
= {\cal P}_d \prod_k u_k \cdot {1 \over ({N\over 2})! }
\biggl[ \sum_{i,j}
a_{i,j} c^\dagger_{i\uparrow}c^\dagger_{j\downarrow}\biggr]^{{N\over 2}}
|0\rangle.
\end{equation}
This form of the wave function is the superposition of
various RVB configurations.  Each valence bond (VB) is represented
by $c^\dagger_{i\uparrow}c^\dagger_{j\downarrow}$, whose amplitude
is $a_{i,j}$.

By investigating the actual form of $a_{i,j}$ from the definition
in eq.~(2.8) for various symmetries,
we can see the difference between the d$_{x^2-y^2}$ wave and
s+$i$d, or extended s wave.
The main difference is that, among the dominant $a_{i,j}$'s,
the next-nearest neighbor bond exactly vanishes
for the d$_{x^2-y^2}$ state, i.e.
\begin{eqnarray}
a_{i,i+x+y}&&= \sum_k \cos(k_x+k_y) \frac{\Delta_k}
{{\varepsilon_k-\mu}+\sqrt{(\varepsilon_k-\mu)^2+\Delta_k^2}}
\nonumber \\
&&=0,
\end{eqnarray}
because of the antisymmetric nature for exchanging $k_x$ and $k_y$.
On the other hand, for the s+$i$d and extended s cases, $a_{i,i+x+y}$
has a comparable amplitude with the nearest-neighbor $a_{i,i+x}$ or
$a_{i,i+y}$.

Due to this main difference, the RVB amplitude for the nearest-neighbor
bond for the d$_{x^2-y^2}$ state is more enhanced
than the s+$i$d or extended s state.
As a result, the d$_{x^2-y^2}$-wave state is the most favorable for
the exchange energy.
(Reversely we can say that the next-nearest-neighbor exchange energy
favors the extended s-wave state.)

To see this more closely, the spin structure factor,
\begin{equation}
S(q)=\frac{1}{N_{\rm a}}\sum_{j,\ell} 4\langle S^z_j S^z_\ell \rangle
e^{iq(r_j-r_\ell)},
\end{equation}
is compared in Fig.~9 for the optimized d$_{x^2-y^2}$ wave and
extended s wave.
The antiferromagnetic correlation is enhanced for the
d$_{x^2-y^2}$ wave, which leads to the gain in the exchange energy.
\par

Next we consider the difference in the kinetic energy in Table I.
The ordered states always lose the kinetic energy
compared with the GWF.
Without taking account of the Gutzwiller projection, we can estimate
\begin{eqnarray}
\langle c^\dagger_{i,\uparrow} c_{j,\uparrow}\rangle
&&={1\over N_a} \sum_k e^{-ik(r_i - r_j)} |v_k|^2
\nonumber \\
&&=-{1\over 4N_a} \sum_k {\epsilon_k - \mu \over E_k} (\cos k_x +\cos k_y).
\end{eqnarray}
In the case of $\Delta=0$, $(\epsilon_k - \mu)/E_k={\rm sign}(\epsilon_k-\mu)$,
but in the presence of $\Delta$, the amplitude of
$\langle c^\dagger_{i,\uparrow} c_{j,\uparrow}\rangle$
generally reduces.
This causes the loss of the kinetic energy for the ordered state.
Since there is a zero point at $k_x = \pm k_y$ on the
Fermi surface for the d$_{x^2-y^2}$-wave state, the loss is the least
among the superconducting states.
This fact also favors the d$_{x^2-y^2}$ symmetry.
\par

Finally we discuss the $\langle -n_in_j/4\rangle$ term.
For example, the GWF gives $\langle -n_in_j/4\rangle = -0.1925$
for $n=0.88$ case.
This is understood from the fact that $\langle n_i\rangle =0.88$, then
$\langle n_i\rangle ^2 /4 = 0.1936$.
As shown in Table I, the value of this term for the other states
does not change so much from that of the GWF.
For example, $\langle -n_in_j/4\rangle$ changes only 0.5\% for
the d$_{x^2-y^2}$ wave.
Actually the charge-density correlation function in the real space
shows that, in the d$_{x^2-y^2}$ case, the amplitude of the nearest
and next-nearest neighbors is slightly enhanced.
This is in contrast to
the GWF which shows dips at those sites due to the exchange hole.
Meanwhile, the extended s wave scarcely modifies the GWF value.
Thus, we expect that an attractive correlation factor gives more
energy gain.
We check the effect of the intersite correlation factor
by introducing eq.~(2.11) to the d$_{x^2-y^2}$-wave state:
${\cal P}_{\rm TLL}\Phi_{\rm SC}$.
However, we have found that the energy is hardly affected by
the intersite correlation.
Therefore the stability of the d$_{x^2-y^2}$ state is mainly caused
by the energy gain of the exchange term which is understood from the
RVB picture as discussed above.
\par

\setcounter{equation}{0}
\section{Phase Diagram}
In this section, we construct a phase diagram in the whole
$J/t$-$n$ plane.
First of all we show the summary of the phase diagram in Fig.~10.
In the following subsections we describe the details of determining
the phase boundaries.

In \S5.1 we discuss the phase boundary between the d$_{x^2-y^2}$-wave
state and the metallic states. 
The stabilization of the extended s wave in the low electron density
is studied in \S5.2.  The properties of the GWF around the
supersymmetric case ($J/t=2$) is discussed in \S5.3.
In \S5.4 the phase separation in the large $J/t$ region is studied.
We consider in \S5.5 the effect of the three-site term
on each phase.
\par\medskip

\noindent
\subsection{Range of d$_{x^2-y^2}$ wave}
To begin with, we determine the phase boundary between the
d$_{x^2-y^2}$-wave state and the metallic state.
Here we use the GWF as the metallic state, because the energy gain
due to the intersite Jastrow
correlation factors is very small.
Figures 11 and 12 show the energy components of
the d$_{x^2-y^2}$-wave function for various electron densities ranging
from $n=1$ down to $0.04$.

The kinetic energy $E_t/t$ shown in Fig.~11 goes to 0 as the electron
density approaches either 0 or 1.
The latter is due to the Gutzwiller projection, ${\cal P}_{\rm d}$.
Furthermore $E_t/t$ is a monotonically increasing function of
$\Delta/t$ in every density.
This means that the kinetic energy does not favor the
superconductivity for every electron density.
On the other hand, $E_J/J$ shown in Fig.~12 decreases at finite
values of $\Delta/t$.  The exchange term gives
the attractive force in the d$_{x^2-y^2}$ channel as expected.
\par

Thus the d$_{x^2-y^2}$ state is not realized without the exchange
interaction ($J/t=0$).
For finite $J/t$, the variational energy
\begin{equation}
E=t\times(E_t/t)+J\times(E_J/J)
\end{equation}
has a minimum at a finite value of $\Delta/t$.
The obtained optimal values of $\Delta/t$ are plotted in Fig.~13.  From
this figure the phase boundary is estimated for every electron density.
For example, $\delta=0.2$ ($n=0.8$) case the phase boundary is
at $J/t=0.1$.
This boundary is shown by a solid line with solid triangles in Fig.~10.
For the higher doping case $n\lsim 0.64$
[Fig.~13(b)], it is difficult to determine the boundary definitely,
because small $\Delta$ seems to be realized for small value of $J/t$.
However, this residual small $\Delta$ is not so meaningful
practically, because the energy gain is as small as that
of the TLL (metallic) state.
Therefore, we estimate the phase boundary by extrapolating
the straight fitting lines as indicated in Fig.~13(b).
The boundary thus determined is represented by
a dashed line with reversed triangles in Fig.~10.
\par

\subsection{Range of extended s wave}
Next we discuss the stability of
the extended s wave in the low density region.
As shown by the analytic treatments in the $n\rightarrow 0$
limit,\cite{Emery,Lin,Kagan,HManousakis}
the free electron state is unstable
against the s-wave bound state for $2<J/t<3.4367$.
We will discuss this possibility of the extended s wave
in the finite electron densities.
\par

In Figs.~14 and 15 we show the expectation values of
energy components for the extended s wave.
[The result of s-wave symmetry (i) of eq.~(2.9) is quantitatively
similar to that of the extended s wave.]
The kinetic energy, $E_t/t$,
rapidly approaches zero as $\Delta/t$ increases.
As regards $E_J/J$, energy gain is hardly seen for small doping as
shown in Fig.~15(b).
This is consistent with the result in the half filling where
$E_t/t$ does not depend on $\Delta$ as discussed before.
On the other hand for small electron densities [Fig.~15(a)], $E_J/J$
appreciably decreases.
This is the cause of the stability of the extended s wave in
the low density.
\par

In Fig.~16 we plot the total energy for $n=0.04$ as an example.
No energy gain from the GWF can be seen for $J/t\le 2$,
while the energy gradually decreases for $J/t>2$.
This is consistent with the analysis of two electron systems, where
the electrons form a s-wave bound state for $J/t \ge 2$.
\par

For finite electron densities,
the stable region of the extended s wave is shown in Fig.~10.
Since the energy gain is very small
and the same order with that obtained by a metallic state
with an intersite correlation factor,
it is not easy to determine accurately the boundary of the
extended s wave.
\par

In Fig.~16 we also plot the energy gain due to the
intersite correlation factor, $\Psi_{\rm TLL}$.
For $J/t<2.0$, the optimal value of $\nu$ in eq.~(2.12) is positive,
which corresponds to the repulsive Jastrow factor.
For example, we get $\nu_{\rm opt}\sim 0.25$ (0.20, 0.14, 0.06) for
$J/t=0$ (0.5, 1.0, 1.5).
Just at $J/t=2.0$ we find that the Jastrow factor does not reduce
the variational energy, which indicates that the GWF is a good
variational function for this ``supersymmetric" case as in 1D.\cite{YO1}
We will come back to this point shortly.
For $J/t>2.0$, $\nu_{\rm opt}$ is negative, namely,
$\nu_{\rm opt}\sim -0.09$ ($-0.19, -0.27$) for $J/t=2.5$
($3.0, 3.5$).
This means that effective attractive interaction exists between
electron.\cite{note2}
\par

\subsection{Supersymmetric Case}
Now we investigate the supersymmetric case ($J/t=2$) more in detail.
For this special parameter, the GWF is shown to be exact in the
low density limit.\cite{YO1}
Actually, the two-electron problem can be solved analytically.
We consider the GWF with two electrons,
\begin{eqnarray}
\Psi_{\rm G}=&&\prod_j[1-n_{j\uparrow}n_{j\downarrow}]\
             c_{k=0\uparrow}^\dagger c_{k=0\downarrow}^\dagger |0\rangle
\nonumber \\
            =&&\sum_{i,j}(1-\delta_{i,j})
             c_{i\uparrow}^\dagger c_{j\downarrow}^\dagger|0\rangle.
\end{eqnarray}
If ${\cal H}$ (including ${\cal H}_3$) is applied to this state,
the resultant equation is written as,
\begin{equation}
{\cal H}\Psi_{\rm G}=-2zt\Psi_{\rm G}+\{2t-J-(z-1)J^{(3)}\}\Xi,
\end{equation}
where $z$ is the number of nearest neighbor sites (here $z=4$) and
\begin{equation}
\Xi=\sum_{i,\tau}c^\dagger_{i+\tau\uparrow}c^\dagger_{i\downarrow}|0\rangle.
\end{equation}
This means that if the condition
\begin{equation}
2t-J-(z-1)J^{(3)}=0
\end{equation}
i.e.
\begin{equation}
\frac{J}{t}=\frac{2}{1+(z-1)J^{(3)}/J}
\end{equation}
is satisfied, the GWF is an exact eigenstate with energy $-2zt$,
irrespective of $N_{\rm a}$
[ in fact $\langle\Psi_{\rm G}|\Xi\rangle/
\langle\Psi_{\rm G}|\Psi_{\rm G}\rangle=O(1/N_{\rm a})$ ].
The energy is identical with the noninteracting case.
As will be mentioned in \S5.5, the free electron state has no
instability against the bound-pair states of both s and d waves
when eq.~(5.6) holds.
Thus the GWF is shown to be the exact ground state.
This property does not depend on the lattice form and dimension.
\par

One can interpret this as the same aspect with the 1D $t$-$J$
model.\cite{YO1}
Namely the electron hopping term, which favors the electrons mutually
apart, and the exchange term, which works attractively, are
well balanced and a kind of ``noninteracting" state is realized.
\par

\subsection{Region of phase separation}
As was shown,\cite{Emery,Putikka1} a phase separation appears
in the 2D $t$-$J$ model for the large $J/t$ region.
Here we discuss it from our variational calculations.
The phase separation occurs between the Heisenberg system
(without hole) and a hole-rich phase.
For the small-$n$ region,
the state with phase separation is approximated as a state with the
Heisenberg island and the remaining vacuum.
The energy of this state is
\begin{equation}
E_{\rm PS}=nE_{\rm Heis.}=-1.1693nJ,
\end{equation}
where $E_{\rm Heis.}$ is the ground state energy of the 2D
Heisenberg model.\cite{TC}
Comparing this energy with the optimized
variational energies of the GWF, the d$_{x^2-y^2}$-wave and
extended s-wave states, we determine the phase boundary,
which is also shown in Fig.~10.
\par

In Fig.~16, the energy $E_{\rm PS}$ is also plotted;
since the $J/t$-dependence is strong, we can determine
the phase boundary accurately in the small electron density region.
Table II summarizes the phase boundary ($J_{\rm c}/t$)
obtained in this way.
The extrapolation of these values to $n=0$ leads us to
$J_{\rm c}/t=3.437$, which is quite close to the exact
value $3.4367$.\cite{HManousakis}
\par

As shown in Fig.~16
the GWF is a good trial function in the low-electron density
and the variational energy has weak $J/t$ dependence.
Therefore we can broadly estimate $J_{\rm c}/t$ as $n\rightarrow 0$
from the GWF.
In the limit of $n\rightarrow 0$ the energy components of the
GWF are analytically expressed as $E_t/t = -4n$,
$E_J/J = -n^2$, $E_3/J^{(3)} = -3n^2$.
These values are the same with the free electron system,
because double occupation can be substantially neglected.
Using $E_{\rm GWF} = -4tn$ and $E_{\rm PS}$ in eq.~(5.7) we obtain
the phase boundary \cite{Emery}
\begin{equation}
J_{\rm c}/t=4/|E_{\rm Heis.}/J|=3.42.
\end{equation}
\par

As $n$ increases, our estimation of the phase boundary becomes
inaccurate.
For one thing, we have assumed that the phase separation takes place
between the Heisenberg island and the vacuum.
Another thing is that we have compared variational energies with
the accurate energy of the phase separation with $E_{\rm Heis}$.
Nevertheless our estimation of the phase boundary agrees well with
other results\cite{Putikka1} up to fairly high density ($n\lsim 0.9$).
\par

\subsection{Effect of the three-site term on the phase diagram}
Next, we consider the effect of the three-site term on the
variational energies and the phase diagram.
In Figs.~17 and 18, $E_3/J^{(3)}$ is shown
for the d$_{x^2-y^2}$-wave and extended s-wave states, respectively.
For the d$_{x^2-y^2}$ case, $E_3/J^{(3)}$ behaves similarly to $E_t/t$,
while for the extended s wave it decreases with
increasing $\Delta/t$ and its shape resembles that of $E_J/J$
rather than $E_t/t$.
\par

First we consider the the boundary between the d$_{x^2-y^2}$
wave and the metallic phase.
In Fig.~19 we plot the $\Delta_{\rm opt}/t$ of the d$_{x^2-y^2}$
wave for some values of $J^{(3)}/J$ and $n=0.8$.
The phase boundary of the d$_{x^2-y^2}$ wave, $J_{\rm c}/t$, where
$\Delta_{\rm opt}$ vanishes, scarcely changes from that
for $J^{(3)}=0$.
This aspect is common for other densities ($n=0.52$-0.96).
Thus, the region of the d$_{x^2-y^2}$ wave hardly changes at least
for $n>0.64$ by the three-site term.
\par

The phase boundary of the extended s wave, on the contrary,
depends on the three-site term considerably.
For the extended s wave, $E_3/J^{(3)}$ decreases with increasing
$\Delta/t$ as shown in Fig.~18.
Thus 
the extended s wave becomes stabler for $J^{(3)}/J>0$
and its region expands.
The shift of the boundary is also shown in Fig.~10 by an arrow
for $E_3/J^{(3)}=0.5$.
Inversely, for $J^{(3)}/J<0$ the energy of the extended s wave increases
so that its region disappears for the case of $J^{(3)}/J=-0.5$,
namely the GWF is stable for every value of $J/t$.
\par

This behavior is consistent with the analysis of two-electron
problem.
If we solve the two electron problem including ${\cal H}_3$,
we obtain the eigenvalue equation
\begin{equation}
\frac{1}{\pi a}K\!\left(\frac{1}{a}\right)=
\frac{J}{4t}(1+3 J^{(3)}/J)
\left[\frac{2}{\pi}K\!\left(\frac{1}{a}\right)-1\right]
\end{equation}
for the two-electron bound state with the extended s symmetry,\cite{note3}
where $a=-E/4t\ge 1$ and $K(k)$ is the complete elliptic
integral of the first kind.
Equation (5.9) has a solution when
\begin{equation}
\frac{J}{4t}(1+3J^{(3)}/J)\ge\frac{1}{2}.
\end{equation}
As a result, for fixed $J^{(3)}/J$, the critical value above which
the two electrons form a bound state is
\begin{equation}
{J_{\rm c} \over t} = {2 \over 1+ 3J^{(3)} / J}.
\end{equation}
This value is exactly the same with the condition eq.~(5.6) where
the GWF becomes an exact eigenstate.
For $J^{(3)}/J=0.5$, this critical value is $J/t = 4/5$.
This value coincides with the boundary shown in Fig.~10
which we determine from the VMC calculations.
When $J^{(3)}/J \le -1/3$, there is no solution for the two-electron
bound state.
\par

The phase boundary to the phase separation is also
estimated.
The shifted boundaries for $J^{(3)}/J=\pm 0.5$ are shown
with open circle and dotted line in Fig.~10.
For $J^{(3)}/J=0.5$
the extended s-wave state becomes
so stable that the phase separation does not take place for small $n$.
On the other hand,
for $J^{(3)}/J=-0.5$ the extended s-wave state is not stabilized
and thus the phase separation expands to smaller $J/t$ region.
\par

\section{Summary and Discussions}
By using the variational Monte Carlo method, we have
studied the $t$-$J$ model with the three-site term
for the square lattice, and obtained a phase diagram
in the full $J/t$-$n$ plane.
We have used Gutzwiller-Jastrow-type wave functions describing
metallic, antiferromagnetic, and various
singlet-pairing superconducting states: the BCS-like s, d$_{xy}$,
s+d-type, and s+$i$d-type waves, which include the d$_{x^2-y^2}$
and extended s waves.
\par
Main results are summarized in the phase diagram shown in Fig.~10.
Remarkable points which we have confirmed and found are:
\par
(1) At half filling, many pairing states including
the d$_{x^2-y^2}$ and the s+$i$d-wave states are degenerate
and the most stable, in accordance with the analysis of the
local SU(2) symmetry.\cite{Affleck,ZGRS}
Nevertheless, the antiferromagnetic state is more stable here.
\par
(2) For plausible parameters of high-$T_{\rm c}$ superconductors,
the d$_{x^2-y^2}$ wave is the most stable state
among various pairing symmetries.
The range of the d$_{x^2-y^2}$ wave hardly changes even if
the three-site term is added.
Substantially, there is no chance to realize s+d, s+$i$d, and
extended s waves in this region.
\par
(3) The d$_{xy}$ wave always has higher energy than the GWF except
for extraordinarily large negative value of $J^{(3)}$.
\par
(4) In the region of low electron density and $J/t>2.0$,
the extended s wave becomes a good state, which almost reproduces
the exact value of the phase boundary for $n\rightarrow 0$.
However the energy gain from the GWF is small.
This phase corresponds to the low-electron-density phase
predicted in the 1D $t$-$J$ model, which consists of two-electron
singlet bound-pairs.\cite{Ogata1d}
\par
(5) Metallic states ${\cal P}\Phi_{\rm F}$ are unstable in itself
against the phase separation for $n\sim 1$ and $J/t>0$.
This is in contrast with the cases of the 1D $t$-$J$, and
1D and 2D Hubbard models.
\par
(6) For the supersymmetric case ($J/t=2$), the GWF is exact
in the $n\rightarrow 0$ limit and behaves like the
noninteracting state in the low-electron-density
region just like the 1D $t$-$J$ model.\cite{YO1}
The GWF is also a very good trial wave function in the region up to
$n\lsim 0.25$.
\par

Our result that the d$_{x^2-y^2}$ wave is stable for the
parameters of high-$T_{\rm c}$ oxides confirms the early
VMC results.\cite{YSS,Gros,GL}
The obtained phase diagram is consistent with the one by
the high-temperature expansion,\cite{Putikka1}
and is qualitatively similar to that of the VMC calculations
by Dagotto \etal\cite{Dagotto}
We suspect, however, that the qualitative and/or quantitative
differences between the previous VMC works\cite{CJZG,Joynt,Dagotto}
and the present one are attributed to some insufficient or erroneous
numerical treatments of the the former works.
\par

We should not consider the results (1)-(5) are completely decisive
because of the approximate trial functions used.
However, the results of the relative stability among various
pairing symmetries are persuasive, because we do not give
partial treatment to a specific symmetry.
\par

We also calculate the variational energies using the Gutzwiller
approximation.\cite{ZGRS}
In contrast to the VMC method, the Gutzwiller approximation
assumes the effect of the projection as some statistical weights.
The details are summarized in Appendix B.
Figure 20 shows the obtained variational energy for various superconducting
states including the d$_{x^2-y^2}$ and extended s waves
for $n\sim 1$ and $J/t=0.5$.
The variational energies for the metallic GWF is also plotted.
This figure is to be compared with Fig.~7 where the same
calculation is carried out using the VMC method.
The Gutzwiller approximation gives a similar result.
In Fig.~20 there is a small region in the vicinity of $n=1$ where
s+$i$d state is realized, which was found before.\cite{Suzu}
This is due to the violation of the
SU(2) symmetry in the approximation.

Figure 21 shows the phase diagram determined from the Gutzwiller
approximation.  As expected from the good coincidence of the
variational energy with the VMC results, the obtained phase diagram
resembles Fig.~10 determined from the VMC method.
Thus, the Gutzwiller approximation gives broadly reasonable
results.
\par

In the following we compare our results with other works.
\par

As for the stability of the d$_{x^2-y^2}$-wave pairing,
our result is common with one for the 2D d-p model.
Using similar variational wave functions for this model,
Asahata and Oguri\cite{OguriAsahata}
recently showed that the d$_{x^2-y^2}$-wave pairing state
appears in the borders of metallic and AF regions in the phase diagram.
On the other hand, concerning the possibility of the extended s wave,
it is pointed out that modification of the Fermi surface structure
in $\Phi$ of eq.~(2.5) is important.

As shown by previous VMC calculations,\cite{YS2tj}
the ferromagnetic region with full momentum corresponding
to Nagaoka's theorem\cite{Nagaoka} exists in the region
$0.6\lsim n<1$ and $J/t\lsim 0.1$.
On the other hand, according to the
high-temperature-expansion,\cite{Putikka2}
partial ferromagnetism appears in the similar region.
In this paper, we have not discussed this ferromagnetism
in detail, since it is confined in a small parameter range.
The relation among the above two calculations and our present results
is an interesting issue left for future studies.
\par

Kagan and Rice\cite{Kagan} showed that the p-wave
pairing is the leading instability for small values of $J/t$ and
low-electron densities by means of the $t$-matrix technique.
In this paper we have not discussed this possibility, because
its energy gain seems extremely small.\cite{HManousakis}
It is rather difficult to detect a tiny energy gain by the
ordinary VMC calculations; moreover such a small energy gain is
easily overcome by introducing the Jastrow correlation factors in
the metallic states.
\par

Another interesting issue is the possibility of spin-charge
separation in the 2D systems.  From the density correlation
function, Putikka \etal\cite{Putikka3}
discussed the spin-charge separation.
However as claimed later, it is not yet conclusive.\cite{Chen}
To study TLL states for the metallic phase, a VMC method\cite{VG}
and a Green function Monte Carlo method\cite{ChenLee} have been
also used.
In this paper we have taken account of this possibility by adopting
a TLL-type correlation factor.
Consequently, the energy gain thereof is very small and it is not
likely that the ground-state becomes a Tomonaga-Luttinger liquid state
in the present variational scheme.
\par

\par\bigskip

\centerline{\bf Acknowledgments}
\par\smallskip
The authors thank H.\ Shiba for useful discussions.
One of the authors (H.Y.) is grateful to J.\ Zittartz,
E.\ M\"uller-Hartmann and A.\ Kl\"umper for discussions
and hospitality at University of Cologne, where a part
of the calculations was carried out.
A part of the computations was done using the facilities
of the Supercomputer Center, Institute for Solid State Physics,
University of Tokyo.
This work is partly supported by Grant-in-Aids for Scientific
Research on Priority Areas, ``Anomalous Metallic States
near the Mott Transition" (07237101) and ``Science of High-$T_{\rm c}$
Superconductivity" and for Encouragement of Young Scientists,
given by the Ministry of Education, Science, Sports and Culture.

\appendix
\section{The GWF for the Hubbard Model}
In this appendix, we show the behavior of $E/t$ as
a function of $n$ ($n\sim 1$) for the 2D Hubbard model:
\begin{equation}
{\cal H}_{\rm hub}=-t\sum_{\langle i,j\rangle\ \sigma}
\left(c^\dagger_{i,\sigma}c_{j,\sigma}
+c^\dagger_{j,\sigma}c_{i,\sigma}\right)
+U\sum_j n_{j\uparrow}n_{j\downarrow},
\end{equation}
with the GWF:
\begin{equation}
\Psi_{\rm G}(g)=\prod_j\left[1-(1-g)n_{j\uparrow}n_{j\downarrow}\right]
\Phi_{\rm F}.
\end{equation}
The optimized results near the half filling is shown in Fig.~22;
$E/t$ behaves linearly as $\delta\rightarrow 0$, although
size dependence becomes a little larger as $n$ approaches 1.
This linear behavior is in sharp contrast with the result
of the $t$-$J$ model
as discussed in \S4.1.
\par

This discordance may be intuitively understood as follows.
By neglecting higher-order terms, ${\cal H}$ of eq.~(2.1)
is a canonical transformation of the Hubbard Hamiltonian:
\begin{equation}
{\cal H}=e^{-iS}{\cal H}_{\rm hub}\ e^{iS}.
\end{equation}
Therefore, the expectation value with respect to the GWF is
written as
\begin{eqnarray}
\langle\Psi_{\rm G}|{\cal H}|\Psi_{\rm G}\rangle
&&=\langle\Psi_{\rm G}|e^{-iS}{\cal H}_{\rm hub}e^{iS}|\Psi_{\rm G}\rangle
\nonumber \\
&&=\langle\tilde\Psi|{\cal H}_{\rm hub}|\tilde\Psi\rangle
\end{eqnarray}
where $\tilde\Psi=e^{iS}\Psi_{\rm G}$.
Consequently, we should adopt modified functions $\tilde\Psi$
for the Hubbard model to obtain a similar behavior.
\par

Nevertheless, in 1D, where the modified wave functions are much
better,\cite{YS3,Otsuka} there exists no conflict
between the two models even within the GWF.
And this GWF-result is consistent with the exact analysis,\cite{OS}
which shows that the Hubbard model is connected to the $t$-$J$
model for every $n$ in the strong coupling limit.
Anyway, it is necessary to confirm by some reliable
method that the $t$-$J$ model is connected to the Hubbard model
also for 2D metallic states.
\par

\appendix
\section{Gutzwiller Approximation}
In this Appendix we summarize the Gutzwiller approximation for
the 2D $t$-$J$ model.
This approximation is extensively studied in connection with the
VMC calculations.\cite{ZGRS}
In order to compare with the present results, we only have to include
the estimation of the $\langle n_i n_j/4\rangle $ term.

In the Gutzwiller approximation, the expectation value
\begin{equation}
\langle \psi_0 | {\cal P}_d c_{i,\sigma}^\dagger c_{j,\sigma} {\cal P}_d
| \psi_0 \rangle ,
\end{equation}
is approximated by\cite{ZGRS}
\begin{equation}
\frac{2\delta}{1+\delta}
\langle \psi_0 | c_{i,\sigma}^\dagger c_{j,\sigma} | \psi_0 \rangle,
\end{equation}
where $\delta$ is the doping rate ($\delta=1-n$).
Similarly the exchange term is given by
\begin{equation}
\langle \psi_0 | {\cal P}_d {\mib S}_{i}\cdot {\mib S}_{j} {\cal P}_d
| \psi_0 \rangle =
\frac{4}{(1+\delta)^2}
\langle \psi_0 | {\mib S}_{i}\cdot {\mib S}_{j}  | \psi_0 \rangle.
\end{equation}

For the $\langle n_i n_j/4\rangle $ term, we introduce the following
approximation,
\begin{eqnarray}
\langle \psi_0 | {\cal P}_d && n_i n_j {\cal P}_d | \psi_0 \rangle
\nonumber \\
&&=\frac{4}{(1+\delta)^2}
\langle \psi_0 |
(1-n_{i\downarrow})n_{i\uparrow} n_{j\uparrow}(1-n_{j\downarrow})
\nonumber \\
&& \qquad\qquad\qquad
+(1-n_{i\downarrow})n_{i\uparrow} n_{j\downarrow}(1-n_{j\uparrow})
\nonumber \\
&& \qquad\qquad\qquad
+(1-n_{i\uparrow})n_{i\downarrow} n_{j\uparrow}(1-n_{j\downarrow})
\nonumber \\
&& \qquad\qquad\qquad
+(1-n_{i\uparrow})n_{i\downarrow} n_{j\downarrow}(1-n_{j\uparrow})
| \psi_0 \rangle
\nonumber \\
&&= \frac{4}{(1+\delta)^2} (1-n_\uparrow) (1-n_\downarrow)
\langle \psi_0 |  n_i n_j | \psi_0 \rangle
\nonumber \\
&&= \langle \psi_0 |  n_i n_j | \psi_0 \rangle.
\end{eqnarray}

Using these approximations, self-consistent equations are given by
\begin{eqnarray}
\Delta_k =&& \frac{1}{N_a} \sum_{k'} \gamma_{k-k'} \frac{\Delta_{k'}}{2E_{k'}},
\\
\xi_k    =&& \frac{1}{\frac{3}{4} g_s J+ \frac{1}{4} J} \biggl\{
-g_t t \gamma_k - \tilde \mu
\nonumber \\
         &&+ \biggl( \frac{3}{4} g_s J - \frac{1}{4} J \biggr)
             \frac{1}{N_a} \sum_{k'} \gamma_{k-k'} \frac{\xi_{k'}}{2E_{k'}}
\biggr\}, \\
\delta  =&& \frac{1}{N_a} \sum_k \frac{\xi_k}{E_k},
\end{eqnarray}
where
\begin{eqnarray}
\gamma_k &&= 2(\cos k_x + \cos k_y ), \qquad
E_k = \sqrt{\xi_k^2 + |\Delta_k|^2 },
\nonumber \\
g_t     =&& \frac{2\delta}{1+\delta},\ \ g_s= \frac{4}{(1+\delta)^2},\ \
\tilde \mu = \mu + \frac{1}{N_a} \frac{\partial \langle{\cal H}\rangle}
{\partial \delta}.
\end{eqnarray}
These self-consistent equations are numerically solved for
the d$_{x^2+y^2}$, s+$i$d and extended s-wave states.


\par
\vfil\eject
\begin{figure}
\caption{
Energy components of the Gutzwiller wave function as a function
of electron density.
Each Symbol represents a system size: solid circle ($L=26$), upward
triangle (20), downward triangle (16), circle (14), square (12)
and diamond (10).
Curves are a guide for the eyes.
}
\label{fig:1}
\end{figure}

\begin{figure}
\caption{
Data of Fig.~1 are re-plotted.
$E_t/t$ and $E_3/J^{(3)}$ are plotted versus $1-n$ in (a), and
$E_J/J$ versus $(1-n)^{0.7}$ in (b).
Symbols have the same meanings with Fig.~1.
}
\label{fig:2}
\end{figure}

\begin{figure}
\caption{
Energy expectation values at half filling for the case
(iii) (s+d-type wave) of eq.~(2.9) for several values of $C$
($-1.0$, $-0.6$, $-0.4$, $0$, $0.4$, $0.6$, $1.0$).
The inset is the magnification of the minimum area around
$\Delta/t=0.57$. Symbols for the value of $C$ are common
in the inset.
The system used is $L=10$, and the sample number is
$3$ - $5\times 10^4$.
The value of Green function Monte Carlo method is
$E/J=-1.1693$.~\cite{TC}
}
\label{fig:4}
\end{figure}

\begin{figure}
\caption{
Energy expectation values at half filling for the case
(iv) (s+$i$d-type wave) for several values of $\theta$
(0, $\pi/12$, $\pi/4$, $\pi/2$, $3\pi/4$, $\pi$).
The inset is the magnification of the minimum area around
$0.5$ - $2$.
Symbols of $\theta$ are common with the inset.
The system used is $L=10$, and the sample number is
$3$ - $5\times 10^4$.
}
\label{fig:5}
\end{figure}

\begin{figure}
\caption{
Expectation values of three energy components for s+$i$d-type wave
and for some value of $\theta$ (0, $\pi/4$, $\pi/2$, $3\pi/4$, $\pi$)
in the same energy scale.
Electron density is chosen at $n=0.88$.
The system used is $L=10$, and the sample number is
$3$ - $5\times 10^4$.
Since there are large statistical fluctuations in $E_J/J$
and $E_3/J^{(3)}$ for small $\theta$ and large $\Delta/t$,
we show the data of two Monte Carlo runs for $\theta=0$.
}
\label{fig:6}
\end{figure}

\begin{figure}
\caption{
Total energy of s+$i$d-type wave for some value of
$\theta$ (0, $\pi/4$, $\pi/2$, $3\pi/4$, $\pi$).
$J^{(3)}/J=0$.
The inset is the magnified figure of the minimum area.
Symbols of $\theta$ are common with the inset.
The system used is $L=10$, and the sample number is
$3$ - $5\times 10^4$
}
\label{fig:7}
\end{figure}

\begin{figure}
\caption{
Comparison of total energies among four variational states
as a function of $n$ near the half filling ($J^{(3)}=0$).
Symbols have the same meaning with Figs.~1 and 2.
The arrow on the right axis represents the accurate value
obtained by the Green function Monte Carlo method.~\cite{TC}
}
\label{fig:3}
\end{figure}

\begin{figure}
\caption{
Total energy of the d$_{x^2-y^2}$-wave state versus $\mu/t$ for some values
of $\Delta/t$.
Note that the scale is small.
The arrow on the horizontal axis indicates the noninteracting
value of $\mu$.
$3\times10^3$ samples are used. $L=10$.
}
\label{fig:10}
\end{figure}

\begin{figure}
\caption{
Comparison of spin structure factors between the GWF,
the optimized d$_{x^2-y^2}$ wave ($\Delta/t=0.55$), and
the extended s wave ($\Delta/t=0.55$).
The path of ${\mib q}$ is
$\Gamma (0,0)\rightarrow X (\pi,0)\rightarrow M (\pi,\pi)\rightarrow \Gamma$.
The system with $L=10$ and $5\times10^4$ samples are used.
}
\label{fig:8}
\end{figure}


\begin{figure}
\caption{
Phase diagram for the ground state of the 2D $t$-$J$-type model.
Basically it represents the case of $J^{(3)}/J=0$.
The shifts due to finite $J^{(3)}/J$ are shown by shadowed
arrows with their values.
The exact phase boundary\cite{HManousakis} between the s-wave
bound state and
the phase separation for $n\rightarrow 0$ is shown by a solid
arrow on the abscissa.
For detailed explanations, see the text.
}
\label{fig:14}
\end{figure}

\begin{figure}
\caption{
Expectation values of transfer energy of the d$_{x^2-y^2}$ wave
for (a) low density ($n=0.04$, 0.12, 0.16, 0.24, 0.32, 0.36, and
0.44) and (b) high density ($n=0.52$, 0.60, 0.64, 0.72, 0.80, 0.88,
and 0.96). $L=10$ and the sample number is
$3$ - $5\times 10^4$.
}
\label{fig:11}
\end{figure}

\begin{figure}
\caption{
Expectation values of exchange energy of the d$_{x^2-y^2}$ wave
for (a) low density ($n=0.04$ - 0.52) and (b) high density
($n=0.60$ - 1.0).
$L=10$.
The sample number is $3$ - $5\times 10^4$.
}
\label{fig:12}
\end{figure}

\begin{figure}
\caption{
Optimized value of $\Delta/t$ for the d$_{x^2-y^2}$ wave as a
function of $J/t$ (a) in the high density ($n=0.64$ - 0.96)
and (b) in the medium density ($n=0.32$ - 0.60).
In (a) we fit the data with curves, and in (b) with straight
lines.
The arrow on the vertical axis in (a) is the value for $n=1.0$.
}
\label{fig:13}
\end{figure}

\begin{figure}
\caption{
Expectation values of transfer energy of the extended s wave
for (a) low density ($n=0.04$ - 0.44) and (b) high density
($n=0.52$ - 0.96). $L=10$ and the sample number is
$3$ - $5\times 10^4$.
}
\label{fig:15}
\end{figure}

\begin{figure}
\caption{
Expectation values of exchange energy of the extended s wave
for (a) low density ($n=0.04$ - 0.52) and (b) high density
($n=0.60$ - 1.0).
$L=10$ and the sample number is $3$ - $5\times 10^4$.
Since for large $\Delta/t$ statistical fluctuations are large,
the results of two trials are depicted for a few values
of $n$.
}
\label{fig:16}
\end{figure}

\begin{figure}
\caption{
Comparison among some variational energies as a function of
$J/t$ for $n=0.04$ for the $t$-$J$ model ($J^{(3)}=0$).
For $J/t\le 2$ the optimal extended s wave is reduced to the GWF.
Note that the scale is small.
$L=10$ and the sample number is $2\times 10^5$.
}
\label{fig:17}
\end{figure}

\begin{figure}
\caption{
Expectation values of three-site term of the d$_{x^2-y^2}$ wave
for (a) low density ($n=0.04$ - $0.52$) and (b) high density
($n=0.60$ - $0.96$). $L=10$ and the sample number is
$3$ - $5\times 10^4$.
}
\label{fig:18}
\end{figure}

\begin{figure}
\caption{
Expectation values of the three-site term of the extended s wave
for (a) low density ($n=0.04$ - 0.52) and (b) high density
($n=0.60$ - 0.96).
$L=10$ and the sample number is 3 - $5\times 10^4$.
For the large value of $\Delta/t$ large statistical fluctuations
are observed, and the results of two trials are depicted for
some values of $n$.
}
\label{fig:19}
\end{figure}

\begin{figure}
\caption{
Optimized value of $\Delta/t$ for some values of $J^{(3)}/J$ as
a function of $J/t$.
Solid lines are guides for eyes.
The wave function has the d$_{x^2-y^2}$-wave pairing symmetry.
}
\label{fig:20}
\end{figure}

\begin{figure}
\caption{
Comparison of total energies among three variational states obtained
in the Gutzwiller approximation
as a function of $n$ near the half filling ($J^{(3)}=0$).
}
\label{fig:ga1}
\end{figure}

\begin{figure}
\caption{
Phase diagram obtained in the Gutzwiller approximation
for the ground state of the 2D $t$-$J$-type model.
}
\label{fig:ga2}
\end{figure}

\begin{figure}
\caption{
Total variational energy for 2D Hubbard model as a function of
electron density near the half filling, with respect to
the Gutzwiller wave function.
Used lattice sizes are indicated by the same symbols
with Fig.~1.
Lines are a guide for the eyes.
$3\times 10^4$ samples are used.
}
\label{fig:21}
\end{figure}

\begin{table}
\caption{Differences of energy components from the
GWF for $J/t=0.5$ in the high electron density.
The values of the GWF are given as the references.
The system with $L=10$ is used.
The last digit for each value includes some error.
}
\begin{center}
\begin{tabular}{cccccc} \hline
$n$  & State & $E_t/t$ & $\langle{\bf S}_i{\bf S}_j\rangle$
& $\displaystyle \langle-\frac{n_in_j}{4}\rangle$ & Total \\
\hline
     & GWF           &       0.0 & $-0.2706$ & $-0.25$   & $-0.5206$ \\
\cline{2-6}
     & TLL           &       0.0 &       0.0 &       0.0 &      0.0  \\
1.00 & AF            &       0.0 & $-0.0500$ &       0.0 & $-0.0500$ \\
     & d$_{x^2-y^2}$ &       0.0 & $-0.0493$ &       0.0 & $-0.0493$ \\
     & Exact         &       0.0 & $-0.0641$ &       0.0 & $-0.0641$ \\
\hline
     & GWF           & $-0.1080$ & $-0.2216$ & $-0.2302$ & $-0.5597$ \\
\cline{2-6}
     & TLL           & $\sim 0 $ & $\sim 0 $ & $\sim 0 $ & $-0.0001$ \\
     & AF            & $+0.0098$ & $-0.0543$ & $-0.0000$ & $-0.0447$ \\
0.96 & d$_{x^2-y^2}$ & $+0.0053$ & $-0.0603$ & $-0.0004$ & $-0.0555$ \\
     & s+$i$d        & $+0.0075$ & $-0.0568$ & $-0.0004$ & $-0.0497$ \\
     & Ext.s         & $+0.0037$ & $+0.0011$ & $-0.0000$ & $+0.0048$ \\
\hline
     & GWF           & $-0.3089$ & $-0.1746$ & $-0.1925$ & $-0.6755$ \\
\cline{2-6}
     & TLL           & $-0.0004$ & $+0.0001$ & $+0.0002$ & $-0.0005$ \\
     & AF            & $+0.0173$ & $-0.0268$ & $-0.0000$ & $-0.0100$ \\
0.88 & d$_{x^2-y^2}$ & $+0.0149$ & $-0.0499$ & $-0.0012$ & $-0.0367$ \\
     & s+$i$d        & $+0.0203$ & $-0.0435$ & $-0.0011$ & $-0.0242$ \\
     & Ext.s         & $+0.0212$ & $+0.0007$ & $-0.0003$ & $+0.0216$ \\
\hline
     & GWF           & $-0.4834$ & $-0.1426$ & $-0.1575$ & $-0.7834$ \\
\cline{2-6}
     & TLL           & $-0.0022$ & $+0.0005$ & $+0.0005$ & $-0.0012$ \\
     & AF            &      0.0  &       0.0 &       0.0 &     0.0   \\
0.80 & d$_{x^2-y^2}$ & $+0.0195$ & $-0.0357$ & $-0.0017$ & $-0.0179$ \\
     & s+$i$d        & $+0.0232$ & $-0.0273$ & $-0.0014$ & $-0.0055$ \\
     & Ext.s         & $+0.0297$ & $-0.0005$ & $-0.0005$ & $+0.0288$ \\
\hline
     & GWF           & $-0.6290$ & $-0.1193$ & $-0.1257$ & $-0.8743$ \\
\cline{2-6}
     & TLL           & $-0.0042$ & $+0.0009$ & $+0.0010$ & $-0.0021$ \\
     & AF            &       0.0 &       0.0 &       0.0 &     0.0   \\
0.72 & d$_{x^2-y^2}$ & $+0.0102$ & $-0.0143$ & $-0.0010$ & $-0.0050$ \\
     & s+$i$d        & $+0.0168$ & $-0.0118$ & $-0.0009$ & $+0.0040$ \\
     & Ext.s         & $+0.0226$ & $-0.0002$ & $-0.0004$ & $+0.0219$ \\
\hline
\end{tabular}
\end{center}
\end{table}

\begin{table}
\caption{
Critical values of $J/t$ of the phase boundary
between homogeneous and separated phases for the low electron
density.
The last digit for each value of $J_{\rm c}/t$ includes some errors.
We represent data for $L\ge 14$.
$J^{(3)}/J=0$.
The exact value for $n=0$ is by Hellberg and Manousakis.\cite{HManousakis}
}
\label{table:2}
\begin{tabular}{@{\hspace{\tabcolsep}\extracolsep{\fill}}lclc} \hline
\quad$n$  & $N/N_{\rm a}$ & \ $J_{\rm c}/t$ & State \\ \hline
0.0       & $2/\infty$ & 3.4367 & Exact   \\ \hline
0.0059    & 4/676      & 3.432  & Ext.s   \\
0.01      & 4/400      & 3.429  & Ext.s   \\
0.0156    & 4/256      & 3.423  & Ext.s   \\
0.0204    & 4/196      & 3.418  & Ext.s   \\
0.03      & 12/400     & 3.389  & Ext.s   \\
0.0612    & 12/196     & 3.345  & TLL     \\
0.0816    & 16/196     & 3.313  & TLL     \\
0.0938    & 24/256     & 3.284  & TLL     \\ \hline
\end{tabular}
\end{table}

\end{document}

%% file: paperfm.tex
%
\def\ie{{\it i.e.}}
\def\etal{{\it et\ al.}}
\newcommand{\lsim}
 {\ \raise.35ex\hbox{$<$}\kern-0.75em\lower.5ex\hbox{$\sim$}\ }
\newcommand{\gsim}
 {\ \raise.35ex\hbox{$>$}\kern-0.75em\lower.5ex\hbox{$\sim$}\ }
%
\def\journal #1#2#3#4{#1 {\bf #2} (#4) #3}
\def\PR{Phys.\ Rev.}
\def\PRB{Phys.\ Rev.\ B}
\def\PRL{Phys.\ Rev.\ Lett.}
\def\PRS{Proc.\ Roy.\ Soc.}
\def\SSC{Solid State Commun.}
\def\APL{Appl.\ Phys.\ Lett.}
\def\IJMP{Int.\ J.\ Mod.\ Phys.}
\def\PLA{Phys.\ Lett.\ A}
\def\IJQC{Int.\ J.\ Quantum\ Chem.}
\def\JAP{J.\ Appl.\ Phys.}
\def\JCP{J.\ Chem.\ Phys.}
\def\JPC{J.\ Phys.\ C}
\def\JPCM{J.\ Phys.\ Cond.\ Mat.}
\def\JJAP{Jpn.\ J.\ Appl.\ Phys.}
\def\JPSJ{J.\ Phys.\ Soc.\ Jpn.}
\def\MPLB{Mod.\ Phys.\ Lett.\ B}
\def\PSCA{Physica}
\def\RMP{Rev.\ Mod.\ Phys.}
\def\PTP{Prog.\ Theor.\ Phys.}
\def\ZP{Z.\ Phys.}
\def\ZPB{Z.\ Phys.\ B}
%
\hyphenation{Coul-omb}
\hyphenation{pho-non}
\hyphenation{pho-nons}
\hyphenation{Phys-ics}
\hyphenation{phys-ics}
\hyphenation{There-by}
\hyphenation{var-i-a-tion-al}
\hyphenation{anti-ferro-mag-net}
\hyphenation{anti-ferro-mag-nets}
\hyphenation{anti-ferro-mag-netism}
\hyphenation{Gutz-wil-ler}

%% file: yojul1.bbl
\begin{thebibliography}{99}
\bibitem{note0} This is different from the 1D case,\cite{YS2tj} where
$E_J/J$ is also linear in $\delta$.
\bibitem{note1} To confirm the $\theta$ dependence, we calculate
each energy component as a function of $\theta$, by fixing
$\Delta/t$ at 0.6, which is the optimal value of the d$_x^2-y^2$ wave for
$J/t=0.5$.
Both $E_t/t$ and $E_J/J$ are monotonically decreasing functions
except for the very vicinity of $\theta=0$.
On the contrary, $E_3/J^{(3)}$ is an monotonically increasing
function of $\theta$, although the increment is relatively small.
\bibitem{note3} For the bound state with d-wave symmetry, we find
$$
J_{\rm c}-J_{\rm c}^{(3)}=\frac{4t}{8/\pi-2}\sim 7.32t.
$$
\end{thebibliography}
